\documentclass[final,5p,times,twocolumn]{elsarticle}
\usepackage{graphicx}
\usepackage{amsmath}
\usepackage{geometry}
\usepackage{amssymb}
\biboptions{compress}

\journal{$\cdots$}

\begin{document}
	
\begin{frontmatter}

\title{Scattering of electromagnetic waves from a graphene-coated thin cylinder of \\ left-handed metamaterial}

\author[label1]{Hamid Pashaeiadl}
\author[label1,label2]{Mahin Naserpour}
\author[label1]{Carlos J. Zapata-Rodr\'{i}guez\corref{cor*}}
\cortext[cor*]{Corresponding author:}
\ead{carlos.zapata@uv.es}
\address[label1]{Department of Optics and Optometry and Vision Science, University of Valencia, Dr. Moliner 50, Burjassot 46100, Spain}
\address[label2]{Physics Department, College of Sciences, Shiraz University, Shiraz 71946-84795, Iran}

\begin{abstract}
	In this paper we explored the scattering behavior of thin cylinders made of LHM and coated by a monoatomic graphene layer.
	A spectral tunability of the resonance peaks is evidenced by altering the chemical potential of the graphene coating, a fact that occurs at any state of polarization of the incident plane wave in opposition to the case of scatterers of dielectric core.
	On the contrary, no invisibility condition can be satisfied for dielectric environments.
	A singular performance is also found for cylinders with permittivity and permeability near zero. 
	Practical implementations of our results can be carried out in sensing and wave manipulation driven by metamaterials.
\end{abstract}


\end{frontmatter}

\section{Introduction}

Fabrication of metal-dielectric structures exhibiting exotic electromagnetic properties ranging from microwaves to visible wavelenghts has recently been enabled by current micro- and nano-technology \cite{Valentine08,Gansel09,Kildishev13,Poddubny13}. 
For terahertz frequencies, homogeneous and isotropic artificial materials have effectively been manufactured showing negative permittivity and permeability simultaneously \cite{Soukoulis06,Shalaev07,Zheludev11}. 
Such terahertz metamaterials were proposed for a great variety of applications including polarizers \cite{Averkov12}, subwavelength imaging \cite{Belov07}, far-field focusing \cite{Hashemi16,Hashemi17}, and cloaking \cite{Chen13b}, to mention a few.

The scattering properties of bodies made of metamaterials including negative-index media have been studied considering different shapes like cylinders \cite{Kuzmiak02,Ruppin04}, tubes \cite{Diaz16d,Diaz16c}, spheres \cite{Garcia08,Miroshnichenko09}, and also as coatings over dielectrics and metals \cite{Sun05,Arslanagic06,Wu07,Wang08b}. 
Tunability of such properties are subject to modification of its internal geometrical configuration, which on the other hand is not always easily accessible.

The advent of graphene technology involving for instance the creation of widely tunable devices in the THz regime by simply incorporating atomically-thin layers is currently in fast development \cite{Ju11,Low14}.
Thus, particles coated by monoatomic films have been proposed for electrochemical sensing, invisible cloaking \cite{Chen11c}, and improved biomedical applications based on the photothermal effect \cite{Lim13}. 
Particularly interesting results a recently published paper concerning a graphene-coated sphere in the case that the sphere is of negative refractive index \cite{Bian17}.
Here we investigate the resonant behavior of subwavelenght left-handed cylinders coated by graphene in the terahertz regime, which exhibit an anisotropic response on the polarization of the incident wave field. 
By means of numerical simulations we will show how alterations of the chemical potential in graphene by simply applying a gate voltage can serve to dynamically control the resonant peaks of the resultant scattering cross section of the Mie scatterers.

\section{Theoretical analysis}

\begin{figure}[tbh]
	\centering
	\includegraphics[width=\linewidth]{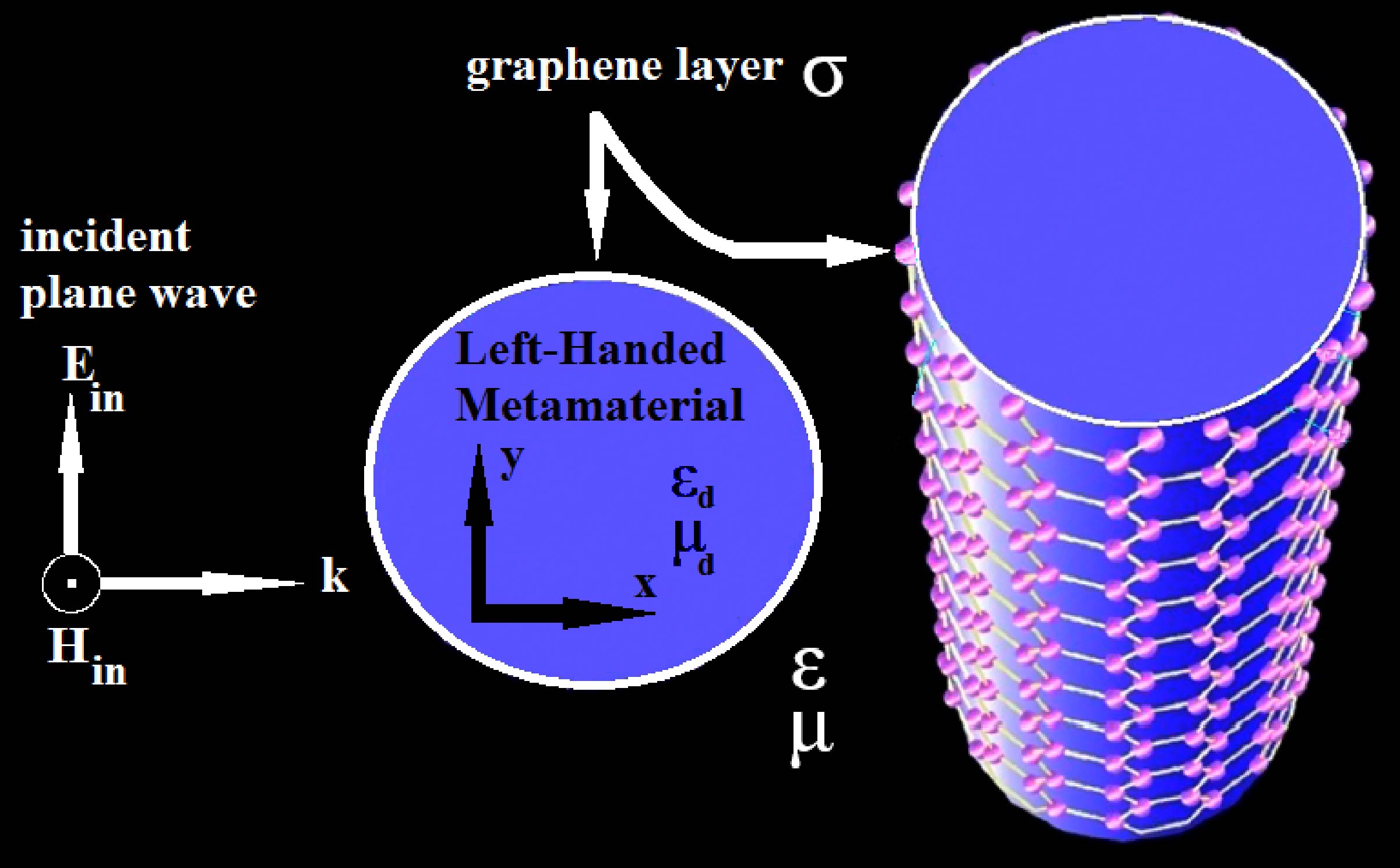}
	\caption{Illustration of the scattering particle coated by an atomically-thin graphene layer and illuminated by a TE$^z$-polarized plane wave.}
	\label{fig01}
\end{figure}

Let us consider a cylindrical dielectric nanowire of radius $R$ and relative permittivity $\epsilon_d $ and magnetic permeability $ \mu_d $ covered by an atomically thin graphene monolayer, and oriented along the $z$ axis, as illustrated in Fig. 1. 
The nanowire is assumed to be placed in an environment medium of the relative permittivity $\epsilon$ and magnetic permeability $\mu$; for simplicity, here we assume that the environment medium is lossless.
The wave vector of the incident radiation is directed along the $x$ axis, and in this case two fundamental polarization configurations can be studied: 1) the electric field vector $\overrightarrow{E_{in}}$ lying in the $xy$ plane (TE$^z$-polarized plane wave or $E_z=0$ modes), and 2) the magnetic field $\overrightarrow{H_{in}}$ lying in the $xy$ plane (TM$^z$-polarized plane wave or $H_z=0$ modes). 
The graphene conductivity was described according to the local random phase approximation of the Kubo formula \cite{Nikitin11} which can be expressed as $\sigma=\sigma_{intra}+\sigma_{inter}$, where the intraband contribution can be written as 
 \begin{equation}
\sigma_{intra}=\dfrac{2ie^2k_BT}{\pi\hslash^2(\omega+i\Gamma)} \ln \left[ 2 \cosh \left( \dfrac{\mu_\mathrm{ch}}{2k_BT} \right) \right],
\end{equation}
and the interband contribution is expressed as
 \begin{eqnarray}
 \sigma_{inter} &=& \dfrac{e^2}{4\hslash} \left[ \dfrac{1}{2} +  \dfrac{1}{\pi} \arctan \left( \dfrac{\hslash\omega-2 \mu_\mathrm{ch}}{2k_BT} \right) \right] \nonumber \\ 
                &-& \dfrac{e^2}{4\hslash} \left[ \dfrac{i}{2\pi} \ln \dfrac{(\hslash\omega + 2 \mu_\mathrm{ch})^2}{(\hslash\omega-2 \mu_\mathrm{ch})^2+(2k_BT)^2} \right].
\end{eqnarray}
In this expression, $-e$ is the charge of an electron, $\hslash$ is the reduced Plank's constant, $k_B$ is Boltzmann's constant, $T$ is the temperature, $\Gamma$ is the charge carriers scattering rate, $\mu_\mathrm{ch}$ is the chemical potential and $\hslash \omega$ is the photon energy. 
Also, an $\exp(-i\omega t)$ time-dependence is implicit throughout the paper, where $\omega$ is the angular frequency, $t$ the time coordinate, and $i=\sqrt{-1}$. 

We followed the Lorenz-Mie scattering method, which is given in detail for instance in Refs. \cite{Shah70,Bussey75}, to analytically estimate the scattering efficiency of the coated nanowire. 
In the case of a TE$^z$-polarized plane wave, as illustrated in Fig.~\ref{fig01}, the electromagnetic field of the incident plane wave may be set as 
 \begin{equation}
\overrightarrow{H_{in}} = \hat{z} H_0 \exp(ikx) = \hat{z} H_0\sum^{+\infty}_{n=-\infty} i^nJ_n(kr) \exp(in\phi) ,
\label{eq03}
\end{equation}
where $ r $ and $ \phi $ are the radial and azimuthal cylindrical coordinates, respectively, $ H_0 $ is a constant amplitude, $ J_n(.) $ is the Bessel function of the first kind and order $ n $, $ k_0=\omega/c $ is the wavenumber in the vacuum, and $ k=k_0\sqrt{\epsilon}\sqrt{\mu} $. 
In Eq.~(\ref{eq03}) we used the Jacobi-Anger expansion of a plane wave in series of cylindrical waves. 
Since
\begin{equation}
\overrightarrow{E_{in}} = \dfrac{\overrightarrow{\bigtriangledown}\times\overrightarrow{H_{in}}}{-i\omega\epsilon\epsilon_0} ,
\end{equation}  
the electric field can be read as
 \begin{equation}
 \overrightarrow{E_{in}} = - E_0 \sum^{+\infty}_{n=-\infty} \left[ \hat{r}i^nn\dfrac{J_n(kr)}{r} + \hat{\phi} k i^{n+1} J^{\prime}_n(kr)] \right]  \exp(in\phi) ,
\end{equation}
where $E_0 = {H_0} / {\omega\epsilon\epsilon_0}$.
Here the prime appearing in $ J^{\prime}_n(\alpha) $ denotes derivative with respect to the variable $ \alpha $. 

The scattered electromagnetic field in the environment medium, $r>R$, may be set as
 \begin{equation}
\overrightarrow{H_{sca}}=\hat{z} H_0 \sum^{+\infty}_{n=-\infty} a_n i^n H^{(1)}_n(kr) \exp(in\phi),
\end{equation}
and
 \begin{equation}
\overrightarrow{E_{sca}}= - E_0 \sum^{+\infty}_{n=-\infty} a_n \left[ \hat{r} i^n\dfrac{n}{r}H^{(1)}_n(kr)+\hat{\phi} k i^{n+1} H^{\prime(1)}_n(kr) \right] \exp(in\phi) ,
\end{equation}
where $ H^{(1)}_n(.)=J_n(.)+iY_n(.)$ is the Hankel function of the first kind and order $ n $, and the coefficients $ a_n $ must be determined. 
The total magnetic field in the environment medium is simply 
\begin{equation}
\overrightarrow{H_{tot}}=\overrightarrow{H_{in}}+\overrightarrow{H_{sca}} .
\end{equation}
Finally, the electromagnetic field in the core of the nanowire $ r<R $ is expressed as
 \begin{equation}
\overrightarrow{H_{d}} = \hat{z} H_0 \sum^{+\infty}_{n=-\infty} b_n i^n J_n(k_dr) \exp(in\phi),
\end{equation}
and
 \begin{equation}
\overrightarrow{E_{d}} = - E_{d0} \sum^{+\infty}_{n=-\infty} b_n \left[ \hat{r} i^n n \dfrac{J_n(k_dr)}{r} + \hat{\phi} k_d i^{n+1} J^{\prime}_n(k_dr) \right] \exp(in\phi)
\end{equation}
where $E_{d0} = {H_0} / {\omega\epsilon_d\epsilon_0}$, the wavenumber $ k_d=k_0\sqrt{\epsilon_d}\sqrt{\mu_d} $. 
Also, the size parameter of the cylinder is defined as the product $k R$ throughout the paper.

The Lorenz-Mie scattering coefficients $a_n$ and $b_n$ are determined by means of the proper boundary conditions. 
Due to the existence of graphene surface conductivity, the boundary conditions at $r=R$ for the case of TE$^z$ polarized waves are
 \begin{equation}
  \hat{\phi} \cdot \overrightarrow{E_d} = \hat{\phi} \cdot (\overrightarrow{E_{sca}} + \overrightarrow{E_{in}})
\end{equation}
and
 \begin{equation}
  \hat{z} \cdot \overrightarrow{H_d} = \hat{z} \cdot (\overrightarrow{H_{sca}}+\overrightarrow{H_{in}}) + \hat{\phi} \cdot \overrightarrow{E_d}\sigma,
\end{equation}
to ensure the continuity of the tangential components of the electromagnetic fields at the graphene layer. 
Since the electric field is set in the $xy$ plane, only the $\phi$-component of the effective conductivity must be taken into account. 
These two equations given above can be set as
 \begin{equation}
  \dfrac{k_d}{\epsilon_d}b_nJ^{\prime}_n(k_dR) = \dfrac{k}{\epsilon} \left[ J^{\prime}_n(kR)+a_nH^{\prime(1)}_n(kR) \right],
\end{equation}
and
 \begin{equation}
  b_n J_n(k_dR) = J_n(kR) + a_nH^{(1)}_n(kR)-\sigma\dfrac{ik_d}{\omega\epsilon_d\epsilon_0}b_nJ^{\prime}_n(k_dR),
\end{equation}
respectively, where the coefficients of different index $n$ can be treated separately due to the linear independence of the multipolar components of the wave field. 
Finally, we obtain the analytical expression of the scattering coefficients
  \begin{eqnarray}
  a_n &=& \left\{ -k \omega \epsilon_d \epsilon_0 J_n(k_dR) J^{\prime}_n(kR) \right.   \nonumber \\
  && \left. + k_d J^{\prime}_n(k_dR) \left[ \omega\epsilon\epsilon_0J_n(kR)-ik\sigma J^{\prime}_n(kR) \right] \right\} /       \nonumber \\
  && \left \{ -k_d\omega\epsilon\epsilon_0H^{(1)}_n(kR)J^{\prime}_n(k_dR) \right.    \nonumber \\
  && \left. + kH^{\prime(1)}_n(kR)[\omega\epsilon_d\epsilon_0J_n(k_dR)+ik_d\sigma J^{\prime}_n(k_dR)] \right \}, \label{eq15}
\end{eqnarray}
and
  \begin{eqnarray}
  b_n &=& k \omega \epsilon_d \epsilon_0 \left[ H^{\prime(1)}_n(kR) J_n(kR) - H^{(1)}_n(kR) J^{\prime}_n(kR) \right]  /   \nonumber \\
  && \left \{ -k_d\omega\epsilon\epsilon_0H^{(1)}_n(kR)J^{\prime}_n(k_dR) \right.   \nonumber \\
  && \left. + kH^{\prime(1)}_n(kR) \left[ \omega\epsilon_d\epsilon_0J_n(k_dR) + i k_d \sigma J^{\prime}_n(k_dR) \right] \right \} .
\end{eqnarray}

As seen above, it is possible to evaluate analytically the scattering coefficients $a_n$ given in Eq.~(\ref{eq15}), which enable the estimation of the scattering efficiency as \cite{Bohren98}
   \begin{equation}
  Q_{sca}=\dfrac{2}{kR}\sum^{-\infty}_{n=+\infty}|a_n|^2.
\end{equation}

\subsection{Equations for TM$^z$ polarziation}

Let us point out that the case of TM$^z$ incident wave fields can be simply determined by means of the duality principle \cite{Balanis89}. 
In this case, we may right
\begin{equation}
 \overrightarrow{E_{in}} = \hat{z} E_0 \exp(ikx) = \hat{z} E_0\sum^{+\infty}_{n=-\infty} i^nJ_n(kr) \exp(in\phi) ,
\end{equation}
for the incident plane wave, 
\begin{equation}
 \overrightarrow{E_{sca}}=\hat{z} E_0 \sum^{+\infty}_{n=-\infty} c_n i^n H^{(1)}_n(kr) \exp(in\phi),
\end{equation}
for the scattered field, and
\begin{equation}
 \overrightarrow{E_{d}} = \hat{z} E_0 \sum^{+\infty}_{n=-\infty} d_n i^n J_n(k_dr) \exp(in\phi),
\end{equation}
for the electric field inside the LHM core.

The boundary conditions at $ r=R $ for TM$^z $-polarized waves reads as
\begin{equation}
\hat{z} \cdot \overrightarrow{E_d}=\hat{z} \cdot (\overrightarrow{E_{sca}}+\overrightarrow{E_{in}}) ,
\end{equation}
and
\begin{eqnarray}
\hat{\phi} \cdot \overrightarrow{H_d}=\hat{\phi} \cdot (\overrightarrow{H_{sca}}+\overrightarrow{H_{in}})-\hat{z} \cdot \overrightarrow{E_d}\sigma .
\end{eqnarray}
Therefore, the analytical expression of the scattering coefficients can be deduced as
\begin{eqnarray}
c_n &=& \left \{ -k_d\mu J^{\prime}_n(k_dR)J_n(kR) \right.     \nonumber \\
     && \left. + \mu_dJ_n(k_dR) \left[ kJ^{\prime}_n(kR) + i\omega\mu\mu_0\sigma J_n(kR) \right] \right \} /    \nonumber \\
     && \left \{-k\mu_dH^{\prime(1)}_n(kR)J_n(k_dR)  \right.    \nonumber \\
     && \left. + \mu H^{(1)}_n(kR) \left[ k_dJ^{\prime}_n(k_dR)-i\omega\mu_0\mu_d\sigma J_n(k_dR) \right] \right\},
\end{eqnarray}
and
\begin{eqnarray}
 d_n &=& \left \{ k\mu_d \left[ H^{(1)}_n(kR)J^{\prime}_n(kR)-H^{\prime(1)}_n(kR)J_n(kR) \right] \right \} /     \nonumber \\
      && \left \{ -k\mu_dH^{\prime(1)}_n(kR)J_n(k_dR) \right.         \nonumber \\
      && \left. + \mu H^{(1)}_n(kR) \left[ k_dJ^{\prime}_n(k_dR)-i\omega\mu_0\mu_d\sigma J_n(k_dR) \right] \right \}.
\end{eqnarray}
Finally, the scattering efficiency can be obtained by using \cite{Bohren98}
\begin{equation}
Q_{sca}=\dfrac{2}{kR}\sum^{-\infty}_{n=+\infty}|c_n|^2.
\end{equation}

\section{Simulations and results}

\subsection{TE$^z$ polarized waves}

\begin{figure}[tbh]
	\centering
	\includegraphics[width=\linewidth]{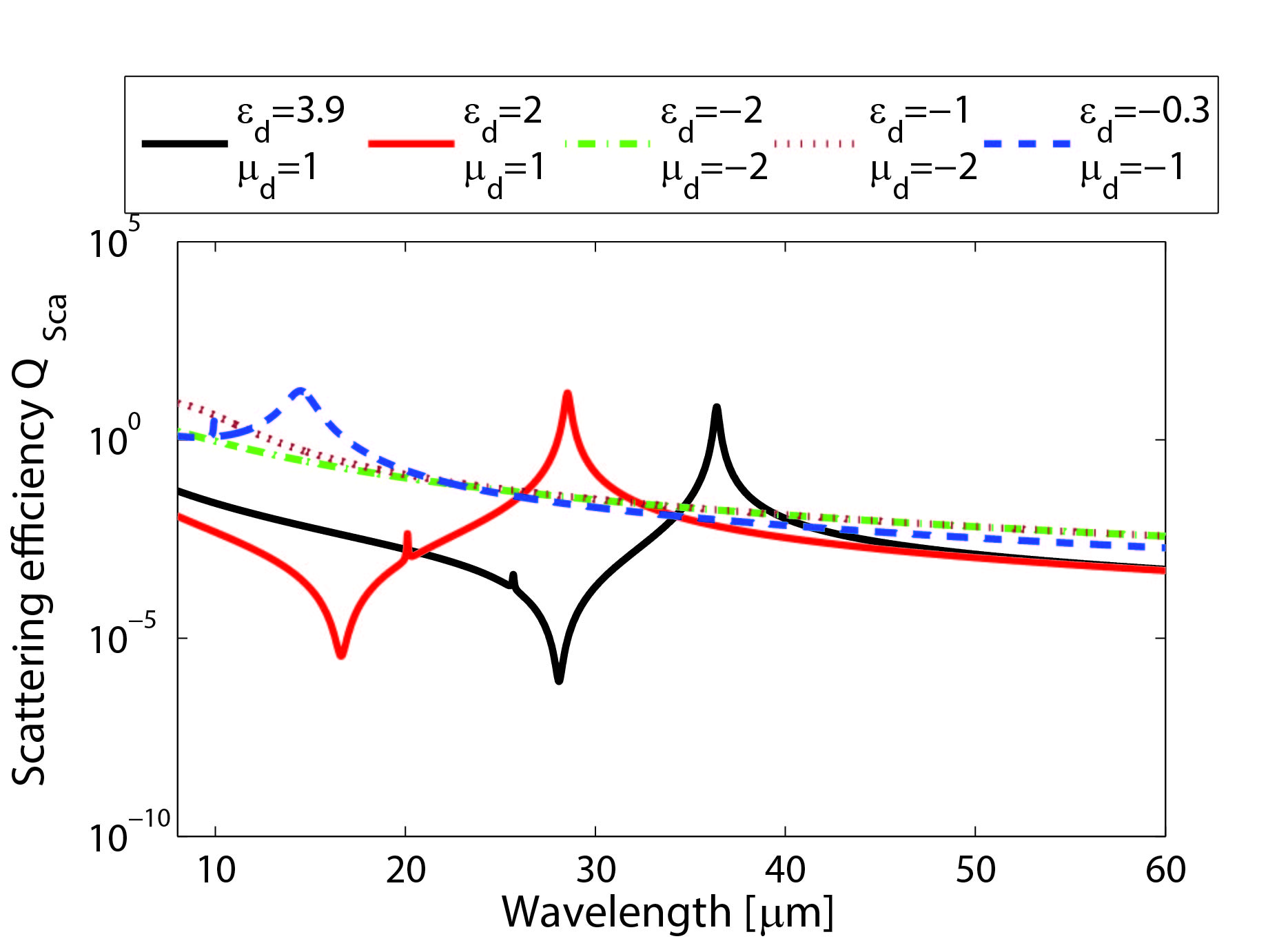}
	\caption{Scattering efficiency spectra, $Q_\mathrm{sca}$ for a uniform-graphene coated cylinder, illuminated by a TE$^z$-polarized plane wave, when $R = 0.5 \mu \mathrm{m}$ and $\mu_\mathrm{ch} = 0.5$eV. 
		In all calculation the charge carriers $\Gamma = 0.1$meV and the temperature $T = 300$K.}
	\label{fig02}
\end{figure}

In Fig.~\ref{fig02} we show the scattering efficiency spectrum of an infinitely-long cylinder illuminated by a TE$^z$-polarized plane wave, that is the electric field lies in the $xy$ plane transversally to the cylinder axis, the latter made of a material of relative permittivity $\epsilon_d$ and permeability $\mu_d$.
In the simulations, $R = 0.5 \mu \mathrm{m}$ is the radius of the cylinder, which in addition is wrapped by a graphene monolayer of chemical potential $\mu_\mathrm{ch} = 0.5$eV. 
As assumed from here on in all calculations, the temperature is $T = 300$K and the charge carriers scattering rate of graphene is $\Gamma = 0.1$meV.
As the most characteristic case in practical implementations, we will consider an environment medium with $\epsilon = 1$ and $\mu = 1$.
For purely-dielectric cylinders, a set of a peak and a minimum in the scattering-efficiency pattern is observed, which are closely located \cite{Riso15}.
Note that the minimum has been proposed as an effective mechanism of invisibility \cite{Chen11,Naserpour17}.
However, when both $\epsilon_d$ and $\mu_d$ are simultaneously negative, as occurs in LHM, the invisibility condition cannot be found.
As a result, a single resonant peak dominates the scattering efficiency spectrum, appearing also non-neglecting secondary peaks related with higher-order resonances.
In general, the full width at half maximum of the main resonances are notably higher for cylinders made of LHM in comparison with scatterers having a dielectric core.
Interestingly, no resonance can be identified for the case $\epsilon_d = -2$ and $\mu_d = -2$.

\begin{figure}[tbh]
	\centering
	\includegraphics[width=\linewidth]{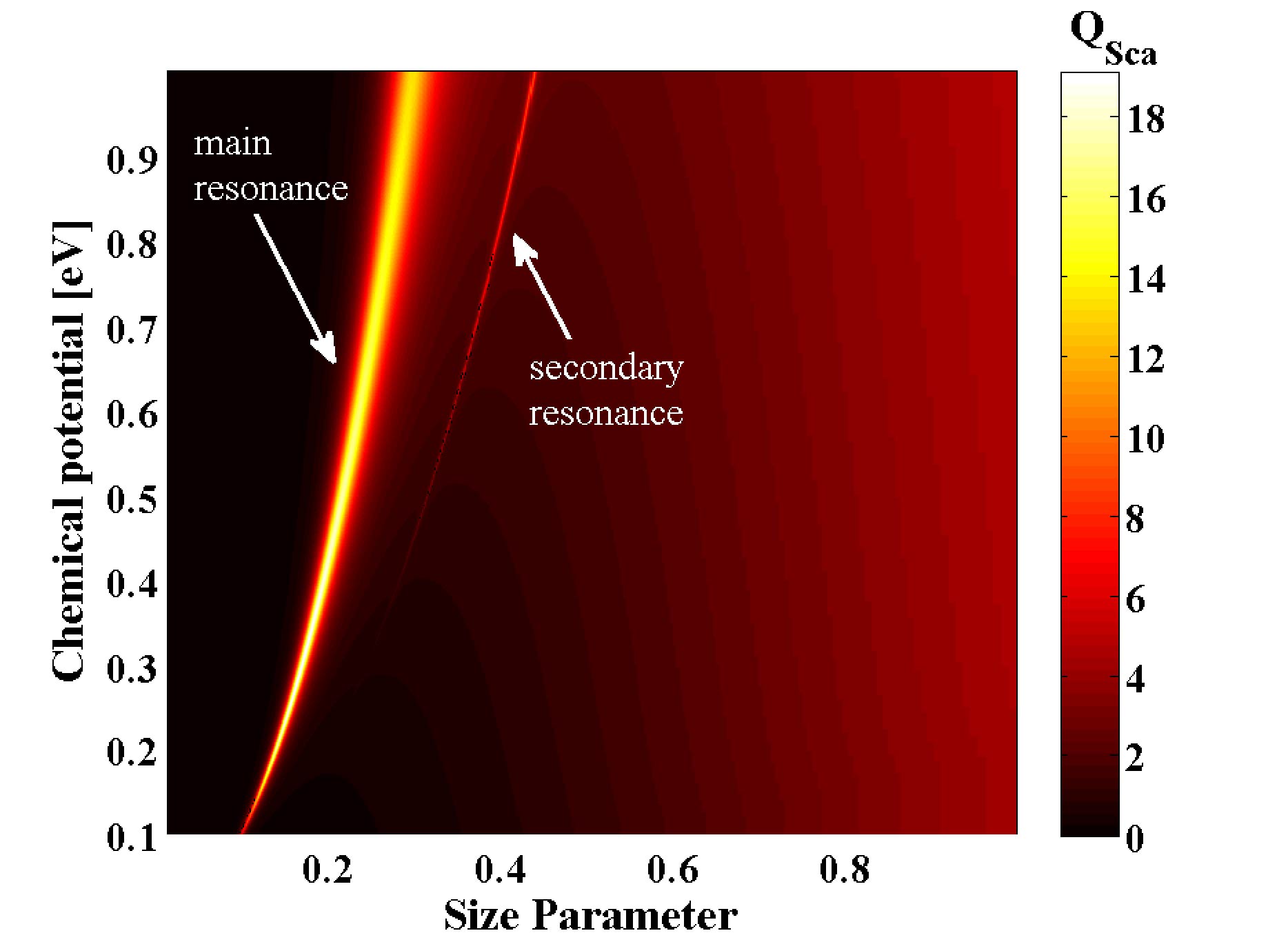}
	\caption{Scattering efficiency spectra for a graphene-coated cylinder when the chemical potential of graphene $\mu_\mathrm{ch}$ varies, maintaining fixed its permittivity $\epsilon_d = -0.3$, permeability $\mu_d = -1$, and core radius $R = 0.5\mu$m.}
	\label{fig03}
\end{figure}

\begin{figure}[tbh]
	\centering
	\includegraphics[width=\linewidth]{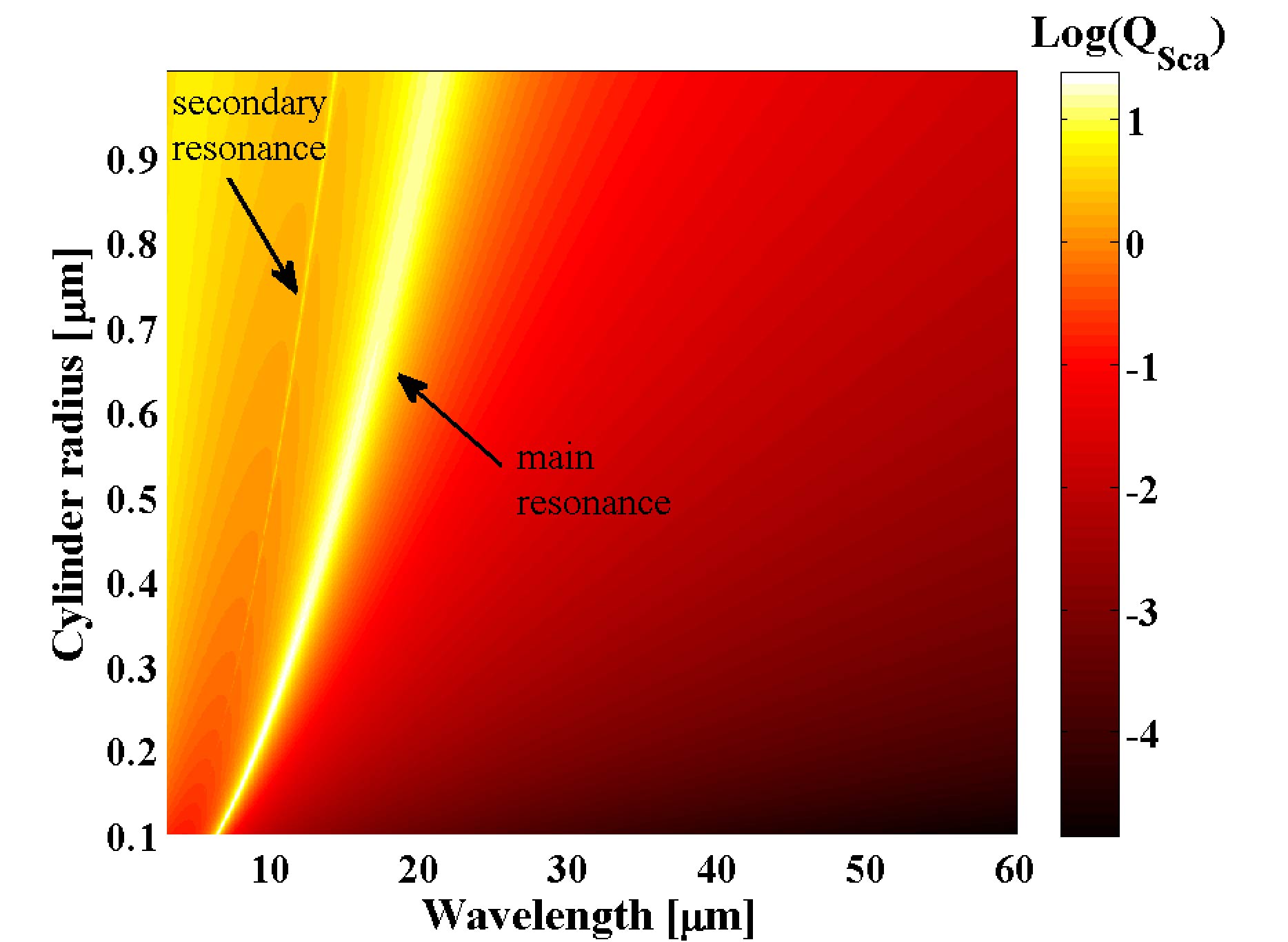}
	\caption{Scattering efficiency spectra for a graphene-coated cylinder when the cylinder radius $R$ varies, maintaining fixed its permittivity $\epsilon_d = -0.3$, permeability $\mu_d = -1$, and chemical potential of graphene $\mu_\mathrm{ch} = 0.5$eV.}
	\label{fig04}
\end{figure}

The resonance peak can be significantly modulated by means of the chemical potential in the graphene coating, an effect that can be achieved for instance by means of an applied gate voltage.
Figure~\ref{fig03} shows the spectrum of the scattering efficiency in the case of a graphene-coated cylinder made of a LHM of $\epsilon_d = -0.3$ and $\mu_d = -1$, varying the chemical potential $\mu_\mathrm{ch}$ from $0.1$eV to $2.0$eV (the latter probably far from realistic setups), in terms of the size paramenter $k R$.
An increasing value of $\mu_\mathrm{ch}$ leads to a significant blue shift of the resonant peak frequency; note that the size parameter is directly proportional to the radiation frequency.
It is worth highlighting that such resonances become extinct in the case of a bare cylinder (not shown in the figure).
The mentioned blue-shift at increasing $\mu_\mathrm{ch}$ occurs not only to the main peak but also to higher-order multipolar resonances.

In order to further analyze the dependence of the resonances upon the cylinder radius, we evaluated the scattering efficiency spectra for a graphene-coated cylinder for a varying parameter $R$, when the core permittivity $\epsilon_d = -0.3$, permeability $\mu_d = -1$, and chemical potential of graphene $\mu_\mathrm{ch} = 0.5$eV are kept fixed. 
In general, the cylinder radius can also modify the frequency of the resonances substantially; a longer radius $R$ displaces the resonant frequency to lower values, affecting not only to the main peak but also to secondary resonances, as shown in Fig.~\ref{fig04}.
This fact has also been observed in graphene-coated dielectric cylinders \cite{Naserpour17}.

\begin{figure}[tbh]
	\centering
	\includegraphics[width=\linewidth]{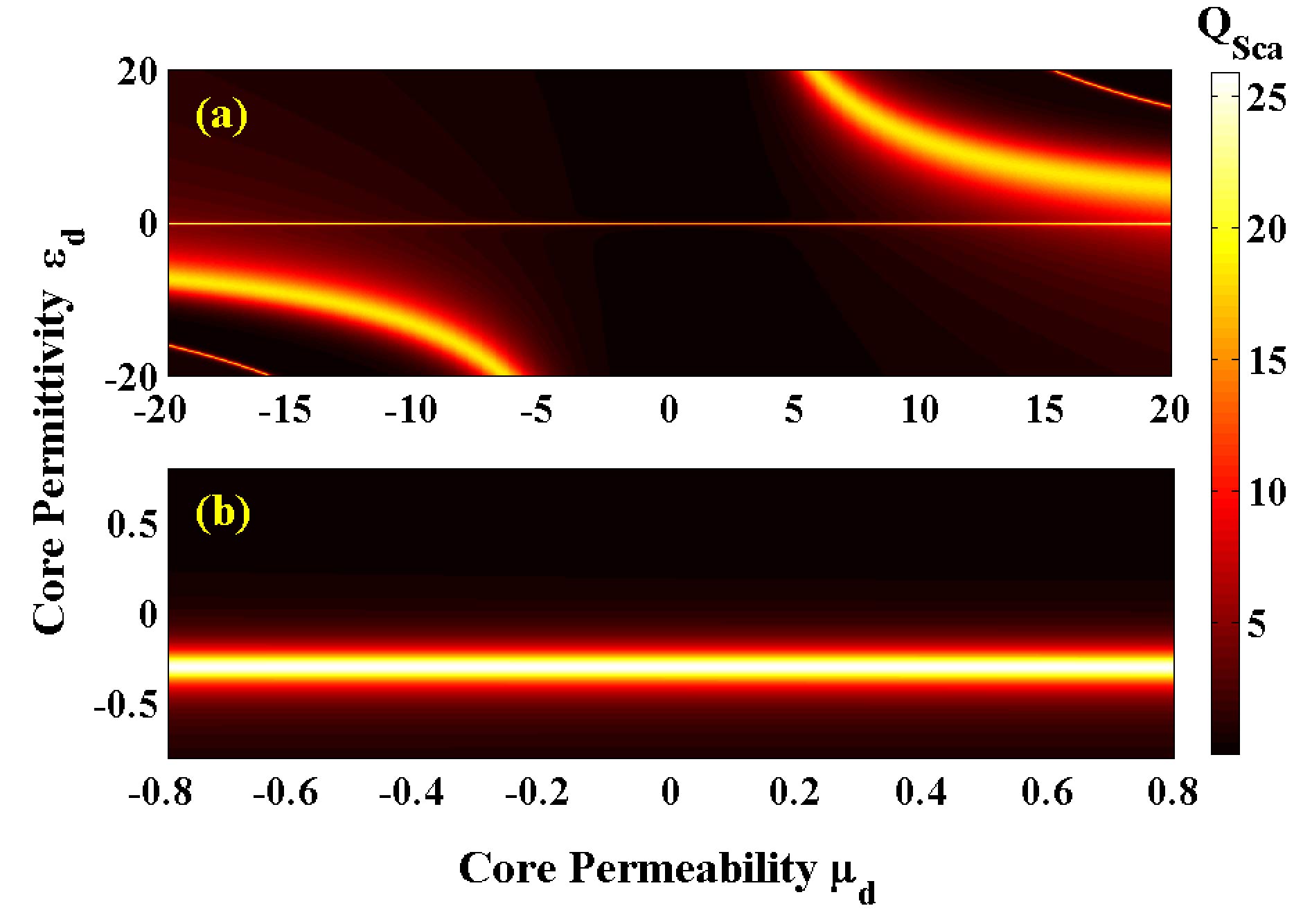}
	\caption{Scattering efficiency for a cylinder coated by a graphene monolayer, and illuminated by a TE$^z$-polarized plane wave, when the electric permittivity $\epsilon_d$ and magnetic permeability $\mu_d$ of the cylinder varies.
		Also, $R = 0.5 \mu \mathrm{m}$, $\mu_\mathrm{ch} = 0.5$eV, and wavelength $ \lambda = 14.5 \mu m$ are maintained fixed.
		In (b) we zoomed (a) in a region of interest.}
	\label{fig05}
\end{figure}

In Fig.~\ref{fig05}(a) we represent the scattering efficiency evaluated in the $\epsilon_d \mu_d$ plane for a cylinder coated by a graphene monolayer, maintaining fixed its radius $R = 0.5 \mu \mathrm{m}$, chemical potential of graphene $\mu_\mathrm{ch} = 0.5$eV, and its wavelength $\lambda = 14.48 \mu m$.
The main peak resonance is nearly mirrored both for dielectrics where $\epsilon_d$ and $\mu_d$ are positive, and for LHM where permittivities and permeabilities remain negative simultaneously.
However, this is no longer true for near-zero values of $\epsilon_d$ and $\mu_d$.
In such case, a resonance arises at $\epsilon_d = -0.3$ independently from the (moderate) value of $\mu_d$, as shown in Fig.~\ref{fig05}(b).

\subsection{TM$^z$ polarized waves}

\begin{figure}[tbh]
	\centering
	\includegraphics[width=\linewidth]{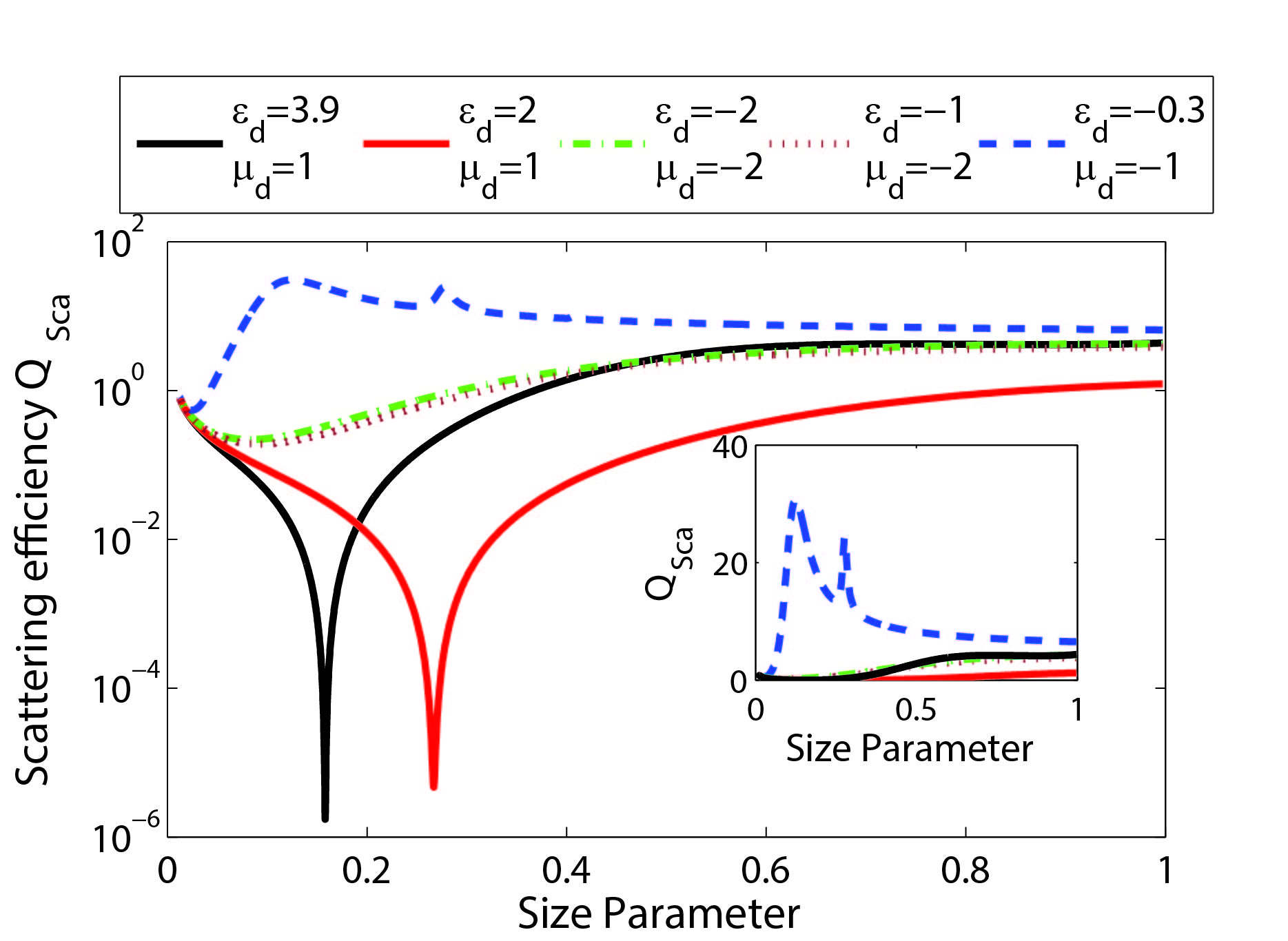}
	\caption{Log-plot of $Q_\mathrm{sca}$ vs. the size parameter $k R$ for a uniform-graphene coated cylinder.
		Again $R = 0.5 \mu \mathrm{m}$ and $\mu_\mathrm{ch} = 0.5$eV.
		Inset: Linear plot of the scattering efficiency.}
	\label{fig06}
\end{figure}

Now let us analyze the scattering behavior of the graphene-coated cylinder when it is illuminated by a TM$^z$-polarized plane wave, that is the electric field is parallel to the cylinder axis.
In Fig.~\ref{fig06} we show $Q_\mathrm{sca}$ for a cylinder made of a material of relative permittivity $\epsilon_d$ and permeability $\mu_d$ taking some different values; in the simulations, its radius is $R = 0.5 \mu \mathrm{m}$ and the chemical potential of graphene is $\mu_\mathrm{ch} = 0.5$eV. 
For such polarization, resonances cannot surge when the cylindrical core is purely dielectric, but a minimum of efficiency features the pattern at low frequencies \cite{Naserpour17}.
On the other hand, some resonance peaks (such that of the lowest frequency located at $\approx 80$THz) are evident for $\epsilon_d = -0.3$ and $\mu_d = -1$.
Such resonances not always emerge in LHM cylinders.
For instance, it does not occur for both near zero (and negative) $\epsilon_d$ and $\mu_d$ when the size parameter remains below the unity.
By the way, the efficiency spectra are now practically identical for near-zero metamaterial cores.
Furthermore, the invisibility condition cannot be found when both $\epsilon_d$ and $\mu_d$ are simultaneously negative.
Finally, note that curves drawn for different radii $R$ are not matching, even when using the size parameter in the representation of $Q_\mathrm{sca}$, due to the dispersive character of the graphene surface conductivity.

\begin{figure}[tbh]
	\centering
	\includegraphics[width=\linewidth]{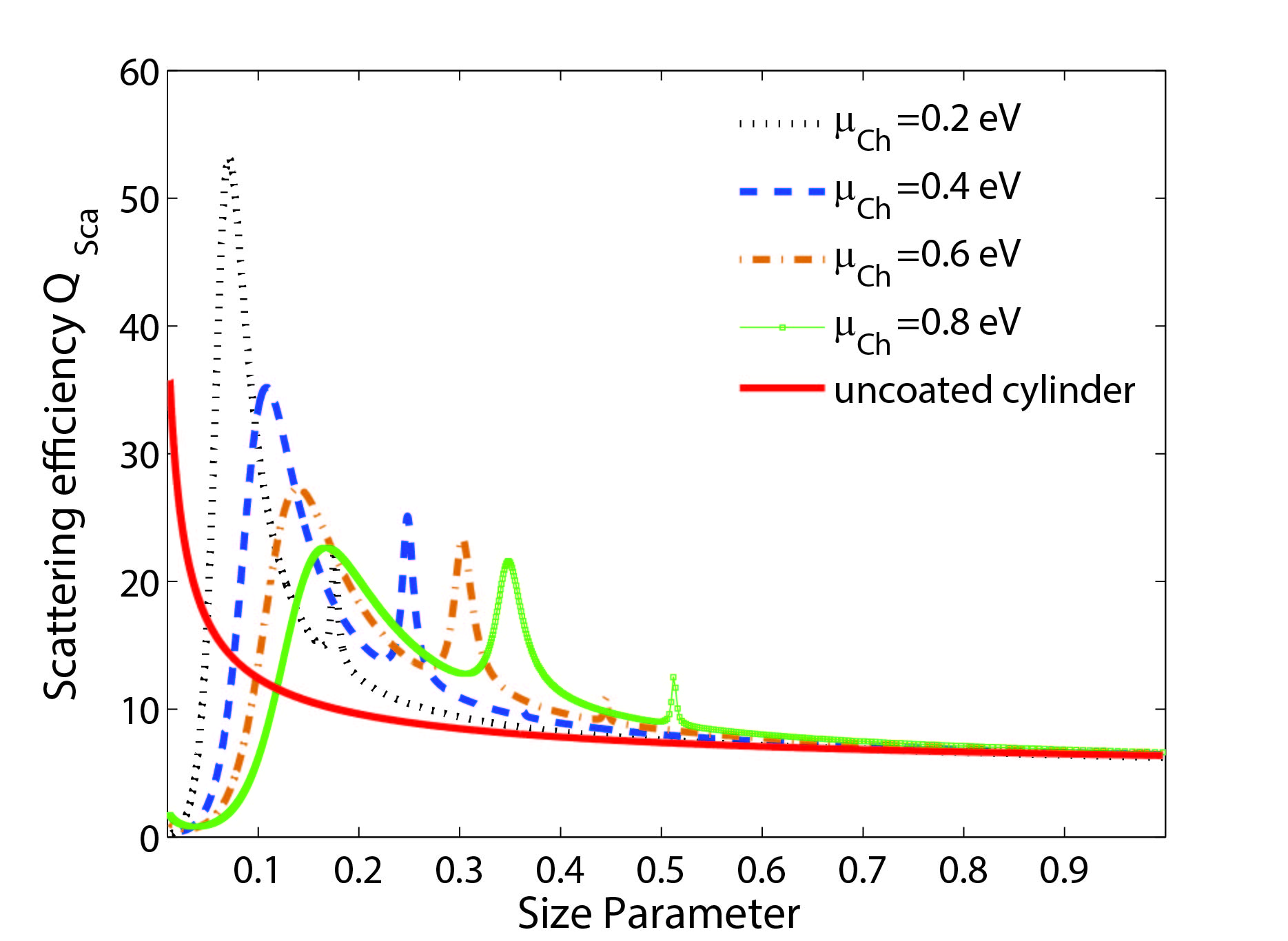}
	\caption{Spectrum of $Q_\mathrm{sca}$ for a graphene-coated cylinder of $\epsilon_d = -0.3$ and $\mu_d = -1$, when the chemical potential $\mu_\mathrm{ch}$ varies, assuming a radius $R = 0.5 \mu \mathrm{m}$.
		Also the case of a bare cylinder is included.}
	\label{fig07}
\end{figure}

Figure~\ref{fig07} shows the spectrum of the scattering efficiency, in terms of the size parameter, in the case of a graphene-coated cylinder made of a LHM of $\epsilon_d = -0.3$ and $\mu_d = -1$, varying the chemical potential $\mu_\mathrm{ch}$ from $0.2$eV to $0.8$eV.
Importantly, the set of multipolar peaks originated by polariton resonances are also strongly governed by the chemical potential of the graphene coating.
Again, higher values of $\mu_\mathrm{ch}$ are associated with higher frequencies of a given resonance.
Finally, a bare cylinder where graphene is not coating the scatterer cannot experience such resonances at so low size parameters. 

\begin{figure}[tbh]
	\centering
	\includegraphics[width=\linewidth]{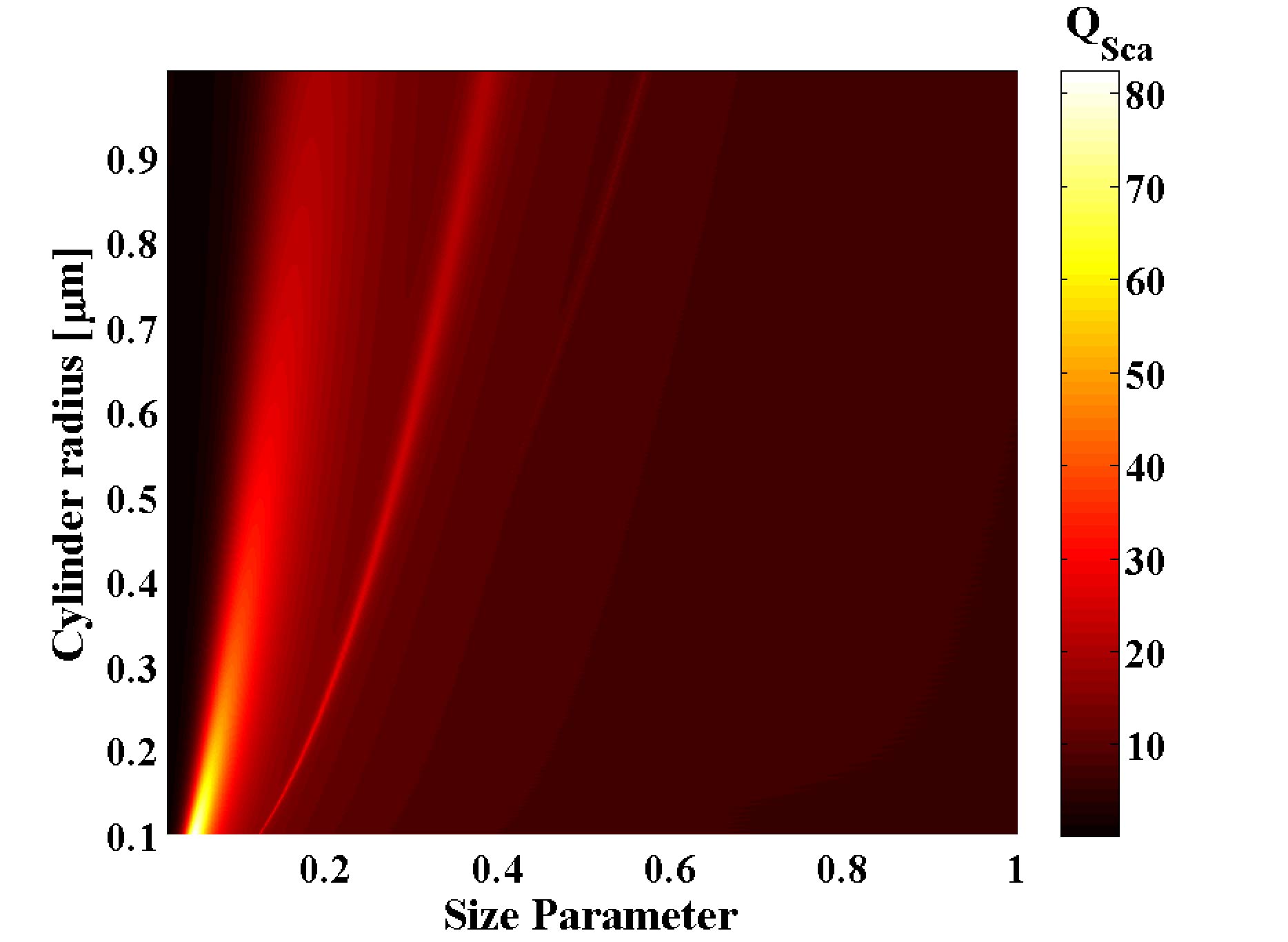}
	\caption{Scattering efficiency spectra for a graphene-coated cylinder when the cylinder radius $R$ varies, maintaining fixed its permittivity $\epsilon_d = -0.3$, permeability $\mu_d = -1$, and chemical potential of graphene $\mu_\mathrm{ch} = 0.5$eV.}
	\label{fig08}
\end{figure}

Here, the cylinder radius also alters the frequency of the resonances significantly, as depicted in Fig.~\ref{fig08} for a graphene-coated cylinder of core permittivity $\epsilon_d = -0.3$, permeability $\mu_d = -1$, and chemical potential of graphene $\mu_\mathrm{ch} = 0.5$eV. 
The longer the radius $R$ of the scatterer the lower frequency where resonances emerge, a fact that involves the main peak and secondary resonances; note that the size parameter itself includes the parameter $R$ and therefore a convenient interpretation of Fig.~\ref{fig08} must be taken carefully.
Note also that the scattering efficiency is notably enhanced for ultra-thin cylinder with a radius below $R = 0.2 \mu$m, a fact that can be difficult to verify in practice.

\begin{figure}[tbh]
	\centering
	\includegraphics[width=\linewidth]{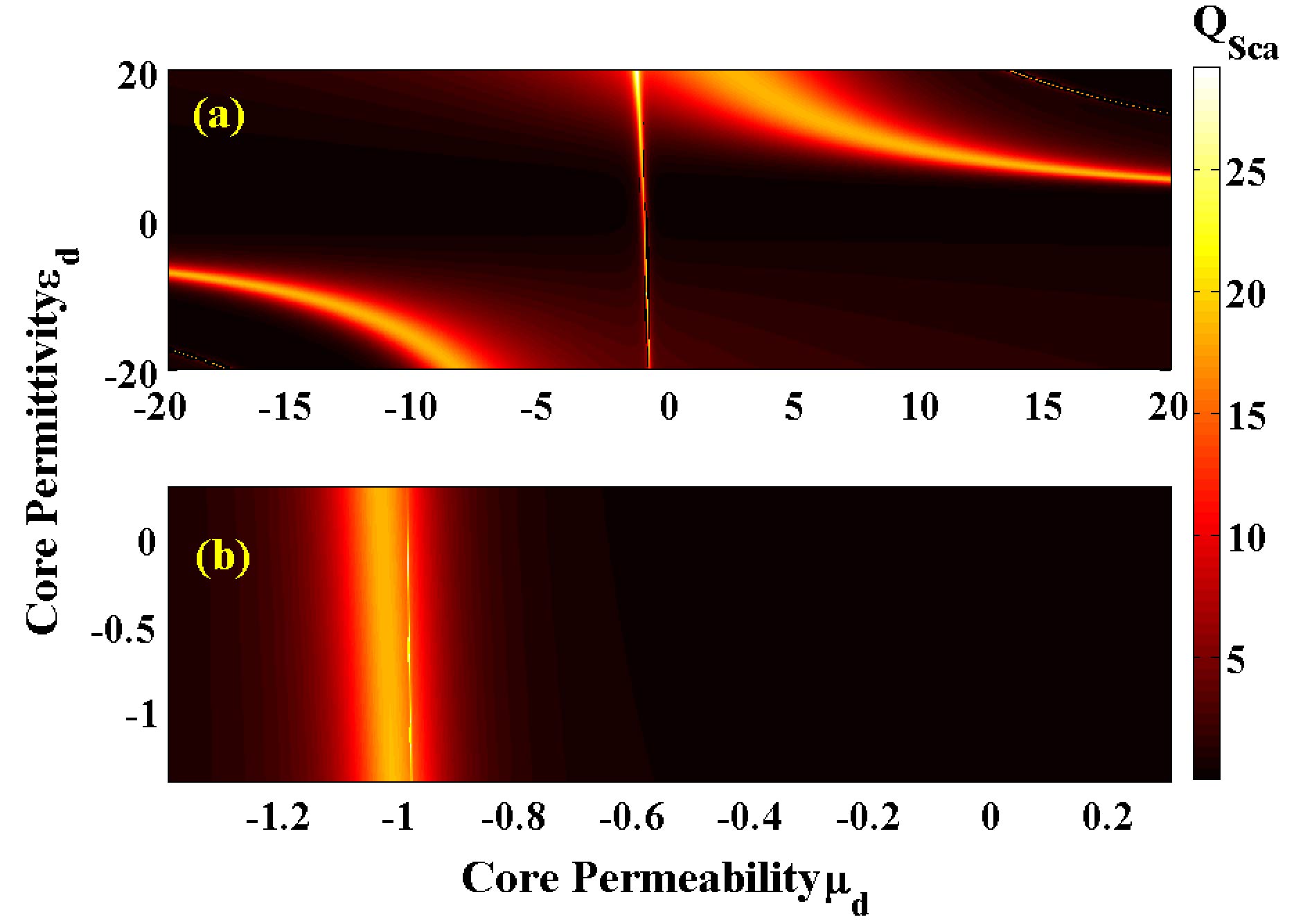}
	\caption{Scattering efficiency of a cylinder coated by a graphene monolayer, and illuminated by a TM$^z$-polarized plane wave, when the electric permittivity $\epsilon_d$ and magnetic permeability $\mu_d$ of the cylinder varies.
		Also, $R = 0.5 \mu \mathrm{m}$, $\mu_\mathrm{ch} = 0.5$eV, and wavelength $= 14.48 \mu m$ are maintained fixed.
		In (b) we zoomed (a) in a region of interest.}
	\label{fig09}
\end{figure}

Finally, the scattering efficiency is determined in the $\epsilon_d \mu_d$ plane for a cylinder coated by graphene, keeping its radius $R = 0.5 \mu \mathrm{m}$, chemical potential of graphene $\mu_\mathrm{ch} = 0.5$eV, and its wavelength ($ = 14.5 \mu m$) as fixed parameters.
The results are depicted in Fig.~\ref{fig09}(a).
As inferred from the duality principle \cite{Balanis89}, we again have the main peak resonance mirrored both in the quadrant for dielectrics and that for LHM.
For near-zero values of $\epsilon_d$ and $\mu_d$, however, now the resonance emerges at values of $\mu_d$ in the vicinities of $\approx -1$, apart from the value of $\epsilon_d$, as evidenced in Fig.~\ref{fig09}(b).

\section{Conclusions}

We investigated the scattering behavior of thin cylinders made of LHM and coated by a monoatomic graphene layer.
Many of the properties we found are evident at any state of polarization (we analyzed TM$^z$ or TE$^z$) of the incident plane wave, as for instance: 
(1) a set of resonant peaks emerge for size parameters below the unity, except maybe for $\epsilon_d$ and $\mu_d$ near-zero. Note that a bare cylinder where graphene is not coating the scatterer cannot experience such resonances at so low size parameters; 
(2) no invisibility condition can be satisfied in opposition to the case of dielectric cores; 
(3) the resonance peaks can be significantly modulated by means of the chemical potential in the graphene coating, an effect that can be achieved for instance by means of an applied gate voltage. In this case, an increasing value of $\mu_\mathrm{ch}$ leads to a significant blue shift of the resonant peak frequencies; 
(4) the radius of the cylinder also alters the frequency of the resonances significantly. 
The longer the radius $R$ of the scatterer the lower frequency where resonances (both main and secondary peaks) emerge.

Finally, we studied the scattering efficiency in the $\epsilon_d \mu_d$ plane.
We revealed that the main peak resonance is nearly mirrored both for dielectrics where $\epsilon_d$ and $\mu_d$ are positive, and for LHM where permittivities and permeabilities remain negative simultaneously.
However, there is an exception in the case of near-zero values of $\epsilon_d$ and $\mu_d$; for TE$^z$-polarized incident plane waves, a main resonance arises for varying values of $\mu_d$ but keeping fixed $\epsilon_d$, however if illumination is carried out by TM$^z$ polarization, the resonance surges at values of nearly-invariant $\mu_d$, apart from the value of $\epsilon_d$.

To conclude, practical implementations of our results can be achieved in sensing and wave manipulation driven by metamaterials.

\section*{Acknowledgments}

This work was supported by the Spanish Ministry of Economy and Competitiveness (MINECO) (TEC2014-53727-C2-1-R).

\bibliographystyle{elsarticle-num}

\begin{thebibliography}{10}
	\expandafter\ifx\csname url\endcsname\relax
	\def\url#1{\texttt{#1}}\fi
	\expandafter\ifx\csname urlprefix\endcsname\relax\def\urlprefix{URL }\fi
	\expandafter\ifx\csname href\endcsname\relax
	\def\href#1#2{#2} \def\path#1{#1}\fi
	
	\bibitem{Valentine08}
	J.~Valentine, S.~Zhang, T.~Zentgraf, E.~Ulin-Avila, D.~A. Genov, G.~Bartal,
	X.~Zhang, Three-dimensional optical metamaterial with a negative refractive
	index, Nature 455~(7211) (2008) 376--379.
	
	\bibitem{Gansel09}
	J.~K. Gansel, M.~Thiel, M.~S. Rill, M.~Decker, K.~Bade, V.~Saile, G.~von
	Freymann, S.~Linden, M.~Wegener, Gold helix photonic metamaterial as
	broadband circular polarizer, Science 325 (2009) 1513--1515.
	
	\bibitem{Kildishev13}
	A.~V. Kildishev, A.~Boltasseva, V.~M. Shalaev, Planar photonics with
	metasurfaces, Science 339~(6125) (2013) 1232009.
	
	\bibitem{Poddubny13}
	A.~Poddubny, I.~Iorsh, P.~Belov, Y.~Kivshar, Hyperbolic metamaterials, Nat.
	Photonics 7 (2013) 948--957.
	
	\bibitem{Soukoulis06}
	C.~M. Soukoulis, M.~Kafesaki, E.~N. Economou, Negative-index materials: New
	frontiers in optics, Adv. Mater. 18 (2006) 1941--1952.
	
	\bibitem{Shalaev07}
	V.~M. Shalaev, Optical negative-index metamaterials, Nat. Photonics 1 (2007)
	41--48.
	
	\bibitem{Zheludev11}
	N.~I. Zheludev, A roadmap for metamaterials, Opt. Photonics News 22 (2011)
	30--35.
	
	\bibitem{Averkov12}
	Y.~O. Averkov, V.~Yakovenko, V.~Yampol’skii, F.~Nori, Conversion of terahertz
	wave polarization at the boundary of a layered superconductor due to the
	resonance excitation of oblique surface waves, Phys. Rev. Lett. 109 (2012)
	027005.
	
	\bibitem{Belov07}
	P.~A. Belov, C.~R. Simovski, P.~Ikonen, M.~G. Silveirinha, Y.~Hao, Image
	transmission with the subwavelength resolution in microwave, terahertz, and
	optical frequency bands, J. Commun. Technol. Electron. 52 (2007) 1009--1022.
	
	\bibitem{Hashemi16}
	M.~Hashemi, A.~Moazami, M.~Naserpour, C.~J. Zapata-Rodr\'{\i}guez, A broadband
	multifocal metalens in the terahertz frequency range, Opt. Commun. 370 (2016)
	306--310.
	
	\bibitem{Hashemi17}
	M.~Hashemi, A.~Moazami, M.~Naserpour, C.~J. Zapata-Rodr{\'\i}guez, Amplitude
	modulation technique for designing metalenses with apodized and enhanced
	resolution focal spots, Opt. Commun. 393 (2017) 77--82.
	
	\bibitem{Chen13b}
	P.-Y. Chen, J.~Soric, Y.~R. Padooru, H.~M. Bernety, A.~B. Yakovlev, A.~Al{\`u},
	Nanostructured graphene metasurface for tunable terahertz cloaking, New J.
	Phys. 15~(12) (2013) 123029.
	
	\bibitem{Kuzmiak02}
	V.~Kuzmiak, A.~A. Maradudin, Scattering properties of a cylinder fabricated
	from a left-handed material, Phys. Rev. B 66~(4) (2002) 045116.
	
	\bibitem{Ruppin04}
	R.~Ruppin, Surface polaritons and extinction properties of a left-handed
	material cylinder, J. Phys. Condens. Matter 16~(34) (2004) 5991.
	
	\bibitem{Diaz16d}
	C.~D\'{\i}az-Avi{\~n}\'o, M.~Naserpour, C.~J. Zapata-Rodr\'{i}guez,
	Optimization of multilayered nanotubes for maximal scattering cancellation,
	Opt. Express 24~(16) (2016) 18184--18196.
	\newblock \href {http://dx.doi.org/10.1364/OE.24.018184}
	{\path{doi:10.1364/OE.24.018184}}.
	
	\bibitem{Diaz16c}
	C.~D\'{\i}az-Avi{\~n}\'o, M.~Naserpour, C.~J. Zapata-Rodr\'{\i}guez, Conditions
	for achieving invisibility of hyperbolic multilayered nanotubes, Opt. Commun.
	381 (2016) 234--239.
	
	\bibitem{Garcia08}
	B.~Garc{\'\i}a-C{\'a}mara, F.~Moreno, F.~Gonz{\'a}lez, J.~Saiz, G.~Videen,
	Light scattering resonances in small particles with electric and magnetic
	properties, J. Opt. Soc. Am. A 25~(2) (2008) 327--334.
	
	\bibitem{Miroshnichenko09}
	A.~E. Miroshnichenko, Non-{Rayleigh} limit of the {Lorenz-Mie} solution and
	suppression of scattering by spheres of negative refractive index, Phys. Rev.
	A 80~(1) (2009) 013808.
	
	\bibitem{Sun05}
	J.~Sun, T.~Jiang, W.~Sun, Y.~Feng, Directive electromagnetic scattering by an
	infinite conducting cylinder coated with left-handed material, in: Antenna
	Technology: Small Antennas and Novel Metamaterials, 2005. IWAT 2005. IEEE
	International Workshop on, IEEE, 2005, pp. 91--94.
	
	\bibitem{Arslanagic06}
	S.~Arslanagic, R.~W. Ziolkowski, O.~Breinbjerg, Excitation of an electrically
	small metamaterial-coated cylinder by an arbitrarily located line source,
	Microw. Opt. Technol. Lett. 48~(12) (2006) 2598--2606.
	
	\bibitem{Wu07}
	Q.~Wu, H.-L. Wang, F.-Y. Meng, L.-W. Li, J.~Wu, Properties of near and far
	fields for the electric line source illumination of a lossless metamaterial
	covered conductor cylinder, Appl. Phys. A 87~(2) (2007) 335--341.
	
	\bibitem{Wang08b}
	H.-L. Wang, Q.~Wu, L.-W. Li, J.~Wu, Properties of near and far fields for the
	electric line source illumination of a lossy metamaterial covered dielectric
	cylinder, Int. J. Infrared Milli. Waves 29~(4) (2008) 373.
	
	\bibitem{Ju11}
	L.~Ju, B.~Geng, J.~Horng, C.~Girit, M.~Martin, Z.~Hao, H.~A. Bechtel, X.~Liang,
	A.~Zettl, Y.~R. Shen, F.~Wang, Graphene plasmonics for tunable terahertz
	metamaterials, Nat. Nanotechnol. 6 (2011) 630--634.
	
	\bibitem{Low14}
	T.~Low, P.~Avouris, Graphene plasmonics for terahertz to mid-infrared
	applications, ACS Nano 8~(2) (2014) 1086.
	
	\bibitem{Chen11c}
	P.-Y. Chen, A.~Al{\`u}, Atomically thin surface cloak using graphene
	monolayers, ACS Nano 5~(7) (2011) 5855--5863.
	
	\bibitem{Lim13}
	D.-K. Lim, A.~Barhoumi, R.~G. Wylie, G.~Reznor, R.~S. Langer, D.~S. Kohane,
	Enhanced photothermal effect of plasmonic nanoparticles coated with reduced
	graphene oxide, Nano Lett. 13~(9) (2013) 4075--4079.
	
	\bibitem{Bian17}
	T.~Bian, X.~Gao, S.~Yu, L.~Jiang, J.~Lu, P.~Leung, Scattering of light from
	graphene-coated nanoparticles of negative refractive index, Optik 136 (2017)
	215--221.
	
	\bibitem{Nikitin11}
	A.~Y. Nikitin, F.~Guinea, F.~Garcia-Vidal, L.~Martin-Moreno, Fields radiated by
	a nanoemitter in a graphene sheet, Phys. Rev. B 84~(19) (2011) 195446.
	
	\bibitem{Shah70}
	G.~A. Shah, Scattering of plane electromagnetic waves by infinite concentric
	circular cylinders at oblique incidence, Mon. Not. R. Astron. Soc. 148 (1970)
	93--102.
	
	\bibitem{Bussey75}
	H.~E. Bussey, J.~H. Richmond, Scattering by a lossy dielectric circular
	cylindrical multilayer, numerical values, IEEE Trans. Antennas Propag. 23
	(1975) 723--725.
	
	\bibitem{Bohren98}
	C.~F. Bohren, D.~R. Huffman, Absorption and scattering of light by small
	particles, Wiley, 1998.
	
	\bibitem{Balanis89}
	C.~A. Balanis, Advanced engineering electromagnetics, Wiley, New York, 1989.
	
	\bibitem{Riso15}
	M.~Riso, M.~Cuevas, R.~A. Depine, Tunable plasmonic enhancement of light
	scattering and absorption in graphene-coated subwavelength wires, J. Opt.
	17~(7) (2015) 075001.
	
	\bibitem{Chen11}
	K.~R. Chen, W.~H. Chu, H.~C. Fang, C.~P. Liu, C.~H. Huang, H.~C. Chui, C.~H.
	Chuang, Y.~L. Lo, C.~Y. Lin, H.~H. Hwung, A.~Y.-G. Fuh, Beyond-limit light
	focusing in the intermediate zone, Opt. Lett. 36 (2011) 4497--4499.
	
	\bibitem{Naserpour17}
	M.~Naserpour, C.~J. Zapata-Rodr{\'\i}guez, S.~M. Vukovi{\'c}, H.~Pashaeiadl,
	M.~R. Beli{\'c}, Tunable invisibility cloaking by using isolated
	graphene-coated nanowires and dimers, Sci. Rep. 7~(1) (2017) 12186.
	
\end{thebibliography}

\end{document}